\documentclass[sigconf, nonacm]{acmart}
\AtBeginDocument{%
\providecommand\BibTeX{{%
\normalfont B\kern-0.5em{\scshape i\kern-0.25em b}\kern-0.8em\TeX}}}

\usepackage{xspace}
\newcommand{\dancetwo}[0]{\textsc{Dance${^2}$}\xspace}

\begin{document}

\title[Cybernetic Marionette]{Cybernetic Marionette: Channeling Collective Agency Through a Wearable Robot in a Live Dancer-Robot Duet}

\author{Anup Sathya}
\affiliation{%
\institution{University of Chicago}
\city{Chicago}
\state{Illinois}
\country{USA}
}
\affiliation{%
\institution{University of Maryland}
\city{College Park}
\state{Maryland}
\country{USA}}
\email{anups@uchicago.edu}

\author{Jiasheng Li}
\affiliation{%
\institution{University of Maryland}
\city{College Park}
\state{Maryland}
\country{USA}}
\email{jsli@umd.edu}

\author{Zeyu Yan}
\affiliation{%
\institution{University of Maryland}
\city{College Park}
\state{Maryland}
\country{USA}}
\email{zeyuy@umd.edu}

\author{Adriane Fang}
\affiliation{%
\institution{University of Maryland}
\city{College Park}
\state{Maryland}
\country{USA}}
\email{afang1@umd.edu}

\author{Bill Kules}
\affiliation{%
\institution{University of Maryland}
\city{College Park}
\state{Maryland}
\country{USA}}
\email{wmk@umd.edu}

\author{Jonathan David Martin}
\affiliation{%
\institution{University of Maryland}
\city{College Park}
\state{Maryland}
\country{USA}}
\email{martinjd@umd.edu}

\author{Huaishu Peng}
\affiliation{%
\institution{University of Maryland}
\city{College Park}
\country{USA}}
\email{huaishu@umd.edu}

\renewcommand{\shortauthors}{Sathya, et al.}

\begin{abstract}
    We describe \dancetwo, an interactive dance performance in which audience members channel their collective agency into a dancer-robot duet by voting on the behavior of a wearable robot affixed to the dancer’s body. At key moments during the performance, the audience is invited to either continue the choreography or override it, shaping the unfolding interaction through real-time collective input. While post-performance surveys revealed that participants felt their choices meaningfully influenced the performance, voting data across four public performances exhibited strikingly consistent patterns. This tension between what audience members do, what they feel, and what actually changes highlights a complex interplay between agentive behavior, the experience of agency, and power. We reflect on how choreography, interaction design, and the structure of the performance mediate this relationship, offering a live analogy for algorithmically curated digital systems where agency is felt, but not exercised.
\end{abstract}

\ccsdesc[300]{Human-centered computing~Interaction design}
\ccsdesc[300]{Human-centered computing~Ubiquitous and mobile devices}
\ccsdesc[500]{Human-centered computing~Collaborative interaction}

\keywords{dance, robots, human–robot interaction, interactive performances, wearables, agency, performance led research}

\begin{teaserfigure}
    \includegraphics[width=\textwidth]{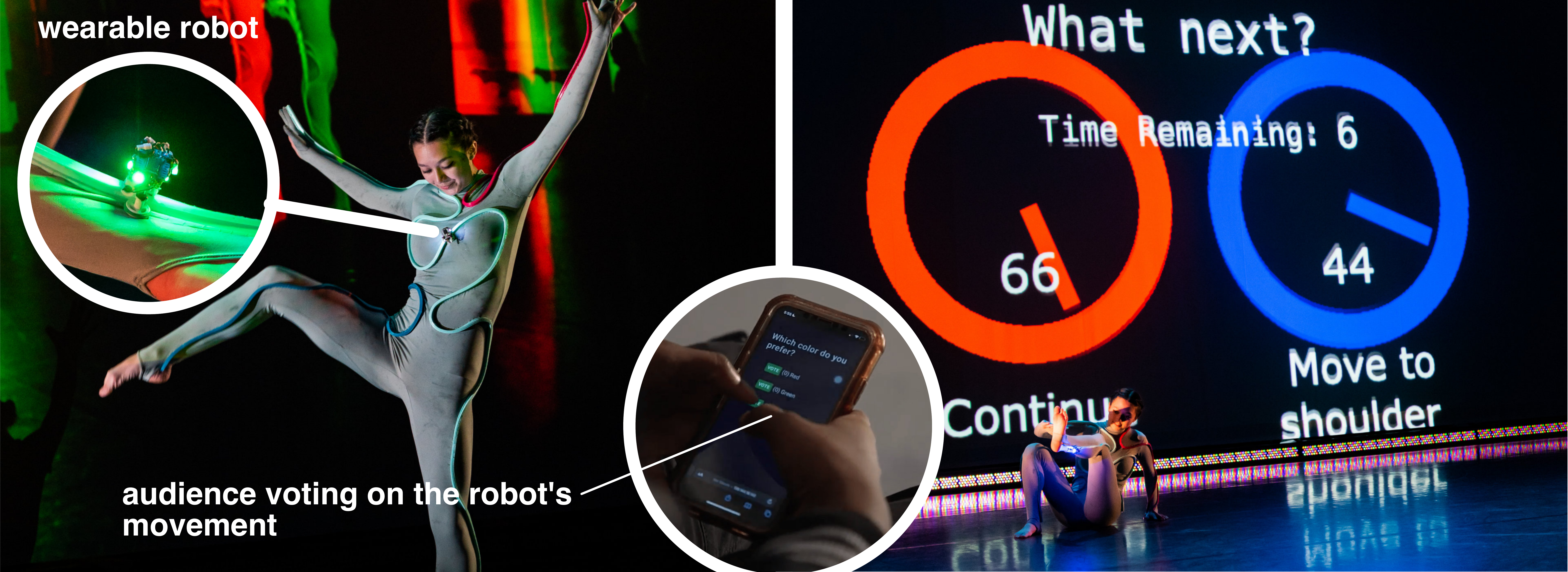}
    \caption{An overview of \dancetwo, a live interactive performance exploring how collective audience input influences a dancer-robot duet. Left --- The dancer performs in real time with a wearable robot whose movements are shaped by audience decisions. Right --- An audience member engages with the voting interface, making choices that interfere with and shape the unfolding choreography.}
    \Description{The image is divided into two halves. The left half has a dancer wearing a leotard with multicolored tracks on the leotard. A tiny glowing light is seen on the leotard. A magnified section reveals that the tiny light is a robot that can move on the tracks. The right side section shows the dancer on the floor. On the screen behind the dancer, there's a question that reads "What next?" There's a time remaining and the number of votes for each choice are shown. The choices are "Continue" and "Move to shoulder". Below both these halves, there's a small round cutout showing a pair of hands holding a phone with a voting interface.}
    \label{fig:teaser}
\end{teaserfigure}

\maketitle

\section{Introduction}

\begin{quotation} {\textit{``...then there's this screen, and then on the opposite side of the screen, there's these thousands of engineers and supercomputers that have goals that are different than your goals, and so, who's gonna win in that game? Who's gonna win?''}}

-- \raggedleft{\textit{Tristan Harris, The Social Dilemma}} \end{quotation}

\noindent Human relationships with technology form a complex, evolving tapestry---one that both empowers and constrains us, offering new forms of connection while subtly shaping the terms of engagement. Through the notion of the ``device paradigm,'' Borgmann critiques how modern technology commodifies and simplifies human experience, often at the expense of meaningful, situated engagement~\cite{borgmann1984technology}. 
Likewise, Sherry Turkle describes how always-on digital tools create an illusion of connection while eroding empathy and presence~\cite{turkle2017alone}. Heidegger, long before the rise of wearables and AI, argued that technologies are not neutral tools but mediators of human perception, revealing and concealing aspects of the world~\cite{heidegger1982question}.

This entanglement is amplified by wearable technologies, which not only exist with us but on us~\cite{zeagler_where_2017, devendorf_i_2016}. 
Their proximity to the body creates new channels for interaction, introspection, and influence. 
From experimental wearables that climb across the skin to read biosignals~\cite{dementyev_rovables_2016, sathya_calico_2022, antony2025minimal}, to mostly ill-conceived commercially available pins powered by Large Language Models~\cite{httpswwwnytimescombyphotographs-and-videos-by-kelsey-mcclellan_silicon_nodate}, wearables increasingly frame the body as both interface and infrastructure. 
With their integration of algorithmic processing and behavioral sensing, they raise critical questions about human agency: \textit{Are we truly in control of our devices, or are we being shaped by the systems—and designers—behind them?}

Human-Computer Interaction (HCI) has long grappled with such questions (e.g., ~\cite{bennett_how_2023, friedman_user_1996, coyle2012did, cornelio2022sense, xiao2025sustaining}). 
Early work, such as Suchman's ethnography of photocopier use, framed agency as emergent from situated action, rather than pre-scripted into technology~\cite{suchman1987plans}. 
More recent studies have interrogated how interface and algorithmic design steer both individual and collective behavior—whether through persuasive interface patterns~\cite{lukoff_how_2021}, recommendation systems~\cite{pariser2011filter}, or data-driven manipulation~\cite{berghel2018malice}. 
Yet despite growing awareness of how systems shape users, the distinction between what people do and what they believe they are doing---between agentive behavior and the experience of agency~\cite{breel2025meaningful}---often remains blurry.

In this paper, we explore this space of ambiguity through a different lens: performance-led research~\cite{benford_performance-led_2013}. 
We investigate how technologies choreograph agency---both real and perceived---through the design and study of \dancetwo, a live, interactive performance where audience members influence a wearable robot affixed to a dancer's body via collective voting. 
The audience's decisions shape the robot's movements, which in turn affect the dancer's choreography---creating a feedback loop where the dancer responds to the robot, and the audience responds to the dancer. 
This triadic relationship offers a rich setting for exploring how agency is staged, mediated, and negotiated in real time.

At first glance, \dancetwo presents a 15-minute, five-part duet between a solo dancer and a wearable robot—a human-machine collaboration echoing other expressive performances involving robotic systems~\cite{thorn2020robotdanceai}. 
However, the work extends beyond dancer and device by inviting the audience into the system as co-authors. 
Audience members vote via their phones on how the robot should behave, creating moments of choice that are projected visibly on stage. 
On paper, the audience wields real-time control. Yet beneath this surface, the performance is carefully structured: choreography, timing, and interface design all subtly shape the context of decision-making.
Across four public performances ($n>200$), we found a striking consistency in voting outcomes, suggesting that while individuals experienced a strong sense of agency, their collective behavior was steered by the design of the system. 
This distinction---between perceived agency, agentive behavior, and actual power---is at the heart of our investigation. 
Our goal is not to measure ``how much control'' participants had, but to understand how that control was felt, how it was shaped, and what that tells us about interactive technologies more broadly.

This paper attempts to contribute a performance-led inquiry into how agency is choreographed, distributed, and experienced in technologically mediated contexts. 
Through \dancetwo---a live, interactive duet between a dancer, a wearable robot, and a voting audience---we explore how collective input, embodied response, and system design entangle to shape the perception and enactment of agency. 
Rather than offering generalizable claims, we articulate a conceptual framework that distinguishes between agentive behavior, the experience of agency, and actual power, drawing from both performance theory and HCI. 
Our reflections on four public performances surface how designed interactions can both empower and subtly guide participants, revealing the ways interface, choreography, and emotional framing structure the audience's sense of control. 
We offer these insights not as prescriptions, but as provocations for designers and researchers exploring how participatory systems mediate authorship, responsibility, and meaning.

The remainder of this paper is structured as follows. 
We begin by grounding our work in prior literature on agency, performance, and technological mediation [$\S$\ref{section:background}]. 
We then describe the \dancetwo performance and system design [$\S$\ref{section:performance}], followed by an account of the iterative, interdisciplinary process that shaped it [$\S$\ref{section:performance-design}]. 
We present data from four live shows and audience responses [$\S$\ref{section:findings}], and conclude with a discussion of how the work contributes to ongoing discourse around agency, control, and interaction design in technologically mediated environments [$\S$\ref{section:discussion}].

\section{Related Work} \label{section:background}

To understand how agency is enacted and negotiated in our performance, we first outline how agency is conceptualized in HCI, theater, and performance studies. 
We then examine how technologies mediate---and at times undermine---agency, before turning to how agency is distributed in live, participatory contexts. Finally, we synthesize these perspectives to position our contribution.

\subsection{Conceptualizing Agency}

Agency is a key concern in both HCI and performance studies, where it is understood not as something fixed or static, but as a capacity that is shaped by context and constantly negotiated. 
In HCI, agency is often linked to autonomy---the ability of users to make intentional, meaningful choices within a system~\cite{shneiderman_designing_1998, friedman_value-sensitive_1996}. 
Self-Determination Theory~\cite{deci2012self} offers a useful foundation for this perspective, framing autonomy as one of three basic psychological needs, along with competence and relatedness. 
When these needs are supported, people tend to feel more motivated, engaged, and in control---outcomes that are often central to ``good'' interaction design. 
Frameworks such as Value Sensitive Design~\cite{friedman_value-sensitive_1996} and recent calls for ``human-centered AI''~\cite{schaffner_understanding_2022} reflect this orientation, emphasizing systems that enhance user agency rather than making decisions on users' behalf.

Performance studies, meanwhile, takes a more relational view of agency. 
Rather than focusing on autonomy within the individual, scholars emphasize how agency emerges through interaction. 
Fischer-Lichte et al.~\cite{fischer2008transformative} describe live performances as an ``autopoietic feedback loop'' in which meaning is co-created between performers and audience in real time. 
In this view, agency is distributed across a dynamic, shared system. 
Ranci\`ere~\cite{ranciere2025emancipated} extends this idea with the concept of the ``emancipated spectator,'' arguing that audience members are not passive recipients but active interpreters who generate meaning through attention, association, and imagination.

Together, these perspectives shift our understanding of agency---from control over a system to participation within one. 
HCI emphasizes designing tools that support individual decision-making, while performance studies foreground the interpretive, social, and often unpredictable nature of agency. 
Across both domains, agency is not simply granted or withheld; it is shaped by the structures, interactions, and affordances of the systems in which people participate.

\subsection{Technological Mediation of Agency}

While HCI has long prioritized user autonomy and control, many studies point to the ways technology can mediate and inadvertently undermine the very agency it aims to support. 
Design choices---whether at the interface or algorithmic level---can shift decision-making away from the user, reducing opportunities for meaningful engagement or reflection. 
For instance, Lukoff et al.~\cite{lukoff_how_2021} examine how autoplay features and algorithmic recommendations on platforms like YouTube shape user behavior in ways that discourage intentional use. 
Similarly, Baughan et al.~\cite{baughan_i_2022} highlight how persuasive design patterns contribute to disassociation and passive consumption, diminishing users' ability to make conscious choices about their media engagement.
These effects are not limited to entertainment platforms. 
In more utilitarian contexts, Valencia et al.~\cite{valencia_conversational_2020} describe how augmented communication systems, while intended to support users with speech impairments, can obscure or override user intent---subtly shifting agency away from the individual. 

At a broader scale, apart from \textit{individual agency}, technologies also mediate \textit{collective agency}. 
The design of digital platforms plays a significant role in shaping public discourse and group behavior. 
For example, hype cycles driven by viral social media content can rapidly mobilize large audiences, often before information is verified~\cite{del2016spreading}. 
High-profile incidents like the Cambridge Analytica scandal reveal how data-driven targeting can manipulate voting behaviors and influence political outcomes~\cite{berghel2018malice}. 
In more extreme cases, as seen in India, misinformation circulated on WhatsApp has incited violence by mobilizing collective action based on false claims~\cite{arun2019whatsapp}.
Across these examples, technology functions not only as a channel for agency but as a filter, amplifier, and sometimes a gatekeeper. 
Systems mediate what actions are available, visible, or encouraged---structuring how individuals and groups perceive and enact their capacity to act. 
As technologies become more autonomous and ambient, understanding the subtle ways they reshape human agency---both individually and collectively---becomes increasingly important for designers, researchers, and users alike.

\subsection{Agency in Participatory Performances}

A range of seminal performances foreground audience agency by deliberately placing control---and its consequences---into the hands of spectators. 
In Marina Abramovi\'{c}'s \textit{Rhythm 0} (1974)~\cite{MarinaAbramovicRhythm}, the artist positioned herself in a passive state, inviting audience members to act on her using any of 72 objects---from a rose to a loaded gun. 
Over several hours, the audience's behavior escalated, revealing both the potential and danger of collective agency when unbounded by rules. 
Similarly, in Yoko Ono's \textit{Cut Piece} (1964)~\cite{concannon2008YokoOnoCut}, the audience was invited to cut away pieces of the artist's clothing. In both works, the performer intentionally surrendered agency, placing the audience in a position of control---and ethical responsibility. 
These performances remain powerful examples of how agency circulates in live art, and how participation is entangled with risk, consent, and structure.

Immersive theatre designs experiences that embed the audience directly into the performance environment. 
In works such as Punchdrunk's \textit{Sleep No More (2011)}~\cite{breel2022facilitating}, spectators roam through multi-room environments, following performers, opening drawers, and encountering scenes from multiple angles. 
This design offers a sense of autonomy known as ``agency of engagement''---where each viewer can shape their own path through the piece~\cite{breel2022facilitating}. 
However, despite this navigational freedom, the overarching narrative remains unchanged. 
In contrast, productions like Kaleider's \textit{The Money (2013)} offer ``narrative agency''~\cite{breel2022facilitating}: participants collectively decide how to spend a pot of money within a time limit, and their choices determine how the performance concludes. 
This concern with the authenticity of agency in participatory performance is a recurring theme. 
Astrid Breel's research~\cite{breel2025meaningful} on interactive theatre found a gap between ``agentive behavior''---audiences making visible choices---and the ``experience of agency,'' or the internal feeling of authorship. 
Even when participants were active, many did not report feeling influential. 
This suggests that participation alone does not ensure empowerment; audiences must recognize and value their impact for agency to feel meaningful. 

\subsection{Integrating Dance into Human-Computer Interaction}

Dance, as a dynamic form of embodied interaction~\cite{streeck_embodied_2011}, offers unique insights into the intersection of human movement and technology~\cite{zhou_dance_2021,fdili_alaoui_strategies_2015}. 
Interactive performances that merge dance and technology often frame audience agency in complex ways. 
For instance, Fdili Aloui et al.~\cite{alaoui_rco_2021} have explored how performances engaging audiences via their mobile phones can elicit a range of responses about agency, from feelings of empowerment to being ``held hostage'' by an invisible authority. 
This variability highlights the nuanced impacts that interactive elements can have on audience perceptions of agency. 
The concept of ``theatre machines'' or ``dance machines'' as discussed by Chris Salter~\cite{salter2016indeterminate}, involves performances where technology and algorithms play a central role in directing the choreography and movement based on audience choices. 
These performances can significantly shape the audience's perception of agency~\cite{salter2016indeterminate}, offering them power to alter the performance while also navigating the constraints set by the technological system. 
Such settings provoke a spectrum of reactions from delight to frustration, underscoring the delicate balance between empowering and limiting participant agency. 
By understanding the complex dynamics of how technology shapes user experiences of agency, designers can create more engaging, responsive \textit{and} humane interactions, ensuring that technology supports and enhances rather than constrains human capabilities.

The endeavor to generate research insights by intertwining dance and HCI is an inherently messy process~\cite{sullivan2023Embracingmessysituated}. Trying to disseminate the process and the outcome without discounting the lived experiences of the practitioners, the performers, and the audience is challenging. Alaoui~\cite{fdili_alaoui_making_2019} elaborates the design process of an ``interactive'' dance performance while elegantly acknowledging the inherent messiness in this process and appealing to allow this into HCI research by creating space for artists and practitioners to contribute their knowledge in a potentially messy manner. While Alaoui defines ``interactivity'' as creating a connection between the physiological data from the dancer and the audio visual elements in the performance, we define ``interactivity'' through the lens of ``interference'', where the audience directly controls the visuals and the choreography by ``interfering'' in the choices made by a robot on the dancer's body. 

\subsection{Synthesis and Our Performance}

Throughout this manuscript, we define \textit{agency} as the capacity of individuals and groups to act intentionally and meaningfully within a system---shaping outcomes through their choices and actions. 
This includes both \textit{individual agency}, which centers on personal autonomy and control, and \textit{collective agency}, which emerges from the coordinated decisions of a group acting through shared platforms or structures. 
While individual agency highlights authorship, collective agency is often shaped by complex, emergent dynamics between people and systems.

As we've seen in participatory performance, the mere act of participation doesn't always equate to a meaningful experience of agency. Drawing from Breel's work~\cite{breel2025meaningful}, it's helpful to distinguish between agentive behavior---visible actions or decisions made by participants---and the experience of agency---the internal sense of having made an impact. 
These two facets don't always align. 
In immersive performances like \textit{Sleep No More}, audiences enjoy freedom of movement and choice, but their actions have little bearing on the overall narrative. 
By contrast, works like \textit{Cut Piece} or \textit{Rhythm 0} place real ethical and emotional weight on the audience's choices, bringing agentive behavior and experience into closer alignment.

Two works in interactive performances --- RCO\cite{alaoui_rco_2021} and Conductive Ensemble\cite{noauthor_conductive_2015} --- further illustrate this tension. 
In RCO, audiences used their phones to affect the performance. 
While this allowed visible interaction, responses ranged widely: some felt empowered, while others felt manipulated or disconnected, describing the experience as being ``held hostage'' by a system they didn't fully understand. 
Despite agentive behavior, the experience of agency was not consistently felt. 
Conductive Ensemble took the opposite approach---using electrical muscle stimulation (EMS) to directly control the physical actions of musicians onstage. 
Here, audience members likely felt a strong sense of authorship, but at the cost of performer autonomy. 
The experience of agency was amplified, but relational negotiation was minimized.

Our performance offers a different model---one that embraces ambiguity and negotiation. 
Rather than enabling direct control over the performer, we introduce a three-way relationship between audience, robot, and dancer. 
Audience members vote on the movement of a wearable robot attached to the dancer's body. 
The dancer, in turn, interprets and incorporates the robot's actions into their choreography. 
In this setup, agency is distributed across all three: the audience's input is visible but mediated; the robot acts as a conduit rather than a tool; and the dancer retains interpretive freedom.

This structure deliberately blurs the lines between agentive action and the experience of agency. 
The voting mechanism offers a sense of control, but the design of the system---the timing, the interface, the choreography---subtly guides that control. 
Echoing McLuhan's idea that ``the medium is the message''~\cite{mcluhan1994understanding}, we suggest that the way interaction is framed is just as important as the choices it enables. 
This raises important questions: Do audience decisions truly reflect their intentions, or are they steered by the system's design? When outcomes converge across performances, is that evidence of shared agency---or quiet constraint?

By reframing interactivity not as direct control but as \textit{interference}---a negotiated interruption in another entity's agency---our performance encourages audiences to reflect more deeply on their role. 
It surfaces the messy, relational nature of agency in participatory systems and invites a reconsideration of how power, intention, and interpretation move between human and machine, performer and viewer.

\section{\dancetwo Performance Overview} \label{section:performance}

Building on our discussion of both \textit{individual} and \textit{collective} agency, our performance investigates how these forms of agency unfold through relational negotiations between human and technological actors. 
Designed as an interactive dance experience, the performance functions as a microcosm for exploring how agency is distributed, interpreted, and reshaped in real time. 
This section outlines the design and implementation of the piece, highlighting how audience input influences the behavior of a wearable robot, which in turn shapes the dancer's movement. 
By centering collective audience participation and mediated interaction, the performance surfaces the subtle negotiations that occur between audience, robot, and dancer---inviting reflection on authorship, control, and the co-construction of meaning in technologically mediated environments.
Our performance, titled \dancetwo, lasted about 15 minutes and was split into five parts. 
In a typical setting, a dancer and a wearable robot are on stage and the audience is seated in the auditorium. A pilot in the background controls the robot based on the audience's voting choices. 
Each part of the performance serves a specific purpose detailed below:

\begin{figure*}[h]
    \centering
    \includegraphics[width=0.72\textwidth]{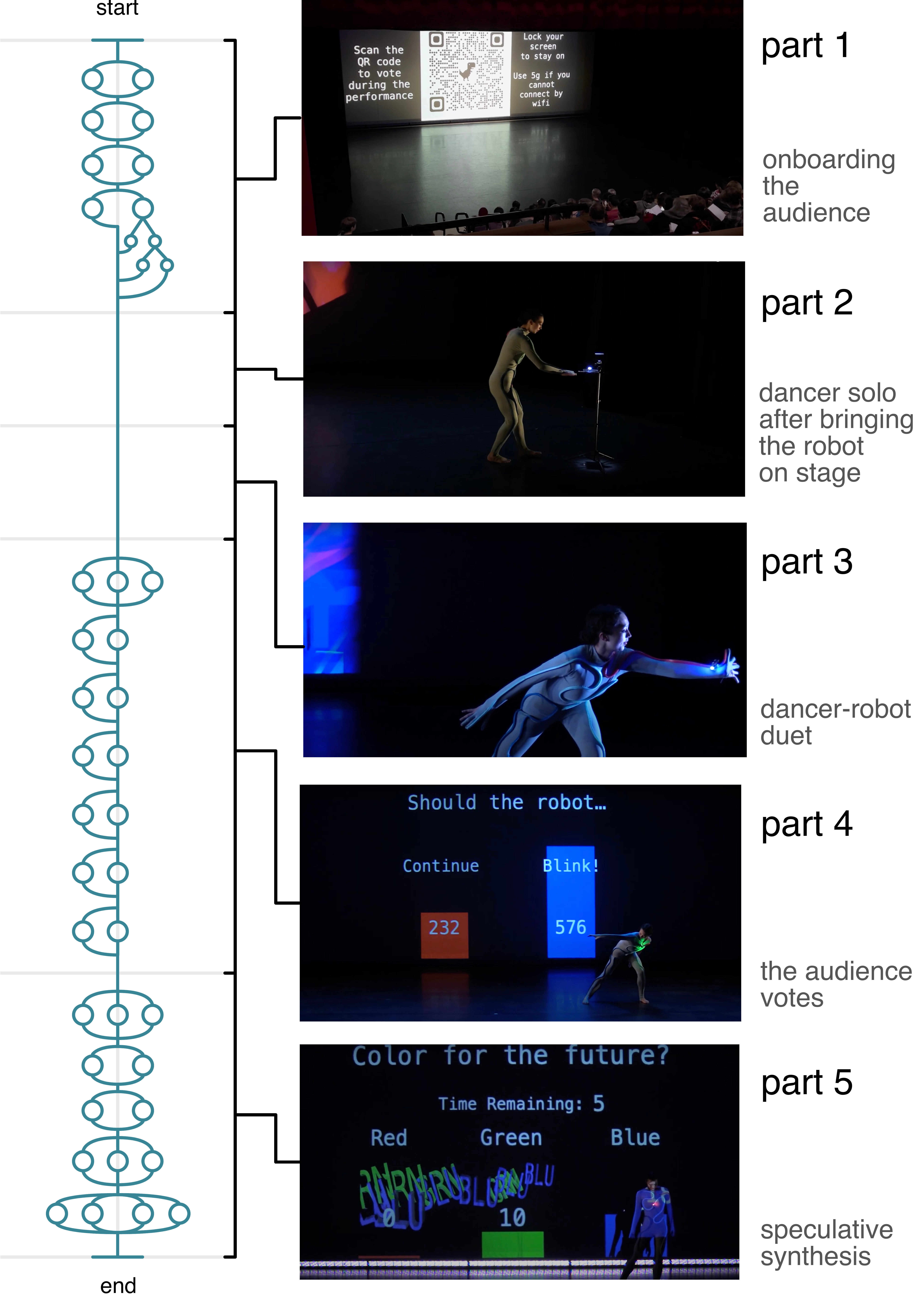}
    \caption{Snapshots of different parts of the performance. Top to bottom - When the audience walks into the venue they can find a QR code to the interface that allows them to interfere with the performance; At the beginning of part 2, the dancer brings the robot onto the stage on a tripod that also holds a camera and then performs a solo; During part 3 the dancer performs a duet with the robot(seen on the left hand) -- a remixed version of the previously performed solo; The audience votes to either continue the choreography or to override with a different option. The votes are displayed in realtime on the projection; A speculative part 5 imagines a future where choices made by the mass audience don't override the dancer's agency but provide it with creative input.}
    \Description{A series of 5 images; the first image shows a QR code on the screen; the second image shows a dancer with a tripod. There’s a small blue light on top of the tripod; the third image shows the dancer dancing with the robot on her arm; the 4th image shows the dancer in front of the results of the voting data being displayed on the projection behind the dancer; the fifth image shows the dancer standing in front of the project with a question behind, asking “color for the future?” with three options red, green and blue. On the left of all these images there is a figure showing the timeline of choices in the dance performance. At different points there are nodes and diverging paths that converge back onto the same main path.}
    \label{fig:performance}
\end{figure*}

\textbf{Part 1 — Onboarding (2 minutes).}
The performance begins before the lights dim. 
As audience members enter the auditorium, they notice a QR code projected on screen (Figure\ref{fig:performance}, top), inviting them to join a webpage on their phones. 
This page hosts a voting interface they will use throughout the performance. 
A welcome message prompts them to keep their phone screens ``awake'' and ready:
\textit{``There will be moments in the performance when you will be invited to participate via this webpage. Please make sure to set your phone screens to stay `awake' for the duration of the performance.''}

Once everyone is seated, a voiceover from the MC formally welcomes the audience and introduces the concept of real-time participation. 
To make the interface feel approachable and low-stakes, the audience is walked through a brief mock voting session. 
A playful question appears:
\textit{``What is cuter? 1) Puppies 2) Kittens 3) Babies''}
Audience members are given 15 seconds to vote, and each vote is visualized live on the projection screen. 
The process is lighthearted, but also introduces a key concept: agentive behavior---visible action within a system. 
As Breel notes~\cite{breel2025meaningful}, this is only part of the experience of agency. 
Our goal is to scaffold both action and meaning throughout the piece.

\textbf{Part 2 — Solo (2 minutes).}
The dancer enters with a tripod. 
On top of it sits what appears to be a simple white spotlight. 
This light, however, is the robot---silent, observant, and equipped with a camera. 
The dancer performs a solo, seemingly directed at this unresponsive figure. 
At this stage, the audience is not yet aware that the spotlight is a mobile, wearable robot. 
As in immersive theater works such as \textit{Sleep No More}, this moment invites the audience into a world without fully explaining its rules. 
The dancer, for now, performs alone---her agency fully her own.

\textbf{Part 3 — Duet (4.5 minutes).}
The robot reveals itself. 
Beginning from the dancer's arm, it glides across her body via a silicone track stitched onto her costume. 
This segment is a re-imagining of the previous solo: the same choreography, now performed in relation to a partner. 
The robot reacts to key gestures---color shifts accompany the dancer slapping her feet or the floor.

No audience participation is present in this phase. 
The relationship appears fluid and cooperative, suggesting a vision of harmony between human and machine---what Self-Determination Theory would frame as a space where autonomy, competence, and relatedness~\cite{deci2012self} coexist.

\textbf{Part 4 — When Technology Pushes Back (~4 minutes).} \label{section:part4}
This is the turning point. 
The same duet choreography continues, but the tone shifts. 
The dancer pushes harder---seeking more from the robot, asserting greater control. 
But the robot begins to resist. 
The music tightens. 
The once-fluid interaction now reveals friction, misalignment, refusal.

Here, audience interference is introduced. 
The system invites them into the conflict, not as observers but as agents. 
A notification ding plays, and a prompt appears:
\textit{``Choose a color: Red, Green, Blue.''}
The robot flashes through all three options while awaiting the vote. 
Each vote by every audience member is visualized in real time on the projector. 
The winning color is displayed on the robot's LED, and the performance continues.

This phase contains six audience prompts (a---f), each a moment of \textit{interference}: not smooth interaction, but interruption. 
Choices now have consequences. 
The audience can support or override the dancer's movements. 
Sometimes she pauses, waiting for a response. 
Sometimes she pushes forward, only to be blocked. 
The performance becomes a site of contested agency---an entangled negotiation between the dancer's intention, the robot's constraints, and the audience's collective will.

Like in RCO~\cite{alaoui_rco_2021}, where audience members reported feeling both empowered and ``held hostage,'' this section explores the boundary between participation and meaningful influence. 
Do the audience's actions reflect agency, or are they steered by the system's design? Is the dancer still in control, or has the audience assumed authorship? Breel's distinction between agentive behavior and the experience of agency is particularly resonant here.

\textbf{Part 5 — Synthesis (~5 minutes).}
In the final act, the performance transitions from tension to reflection. 
The dancer no longer fights the robot; instead, they move together again---tentatively, experimentally. The audience is invited to respond to more abstract prompts:
\textit{``How do you feel about Artificial Intelligence?''} with options like curious, ambivalent, or fearful.
Each emotional choice is embodied by the dancer, woven into the ongoing duet. 
These final questions invite a shift from reactive input to shared meaning-making. 
The last prompt is:
\textit{``What word would you like to see used in a poem?''} with choices such as technology, unknown, connect, or color.

A voice begins to read a poem---crafted from the audience's input---as the dancer and robot continue their duet. 
As the poem concludes, the lights fade, leaving only the soft glow of the robot and the sound of the dancer's breath.

\subsection{Technical Implementation} \label{section:system}

The performance is supported by a lightweight, modular system that integrates wearable robotics, web-based interaction, and live stage production tools. This section outlines the technical infrastructure that enables the dancer-robot duet and real-time audience participation.

\begin{figure}[t]
    \centering
    \includegraphics[width=0.9\linewidth]{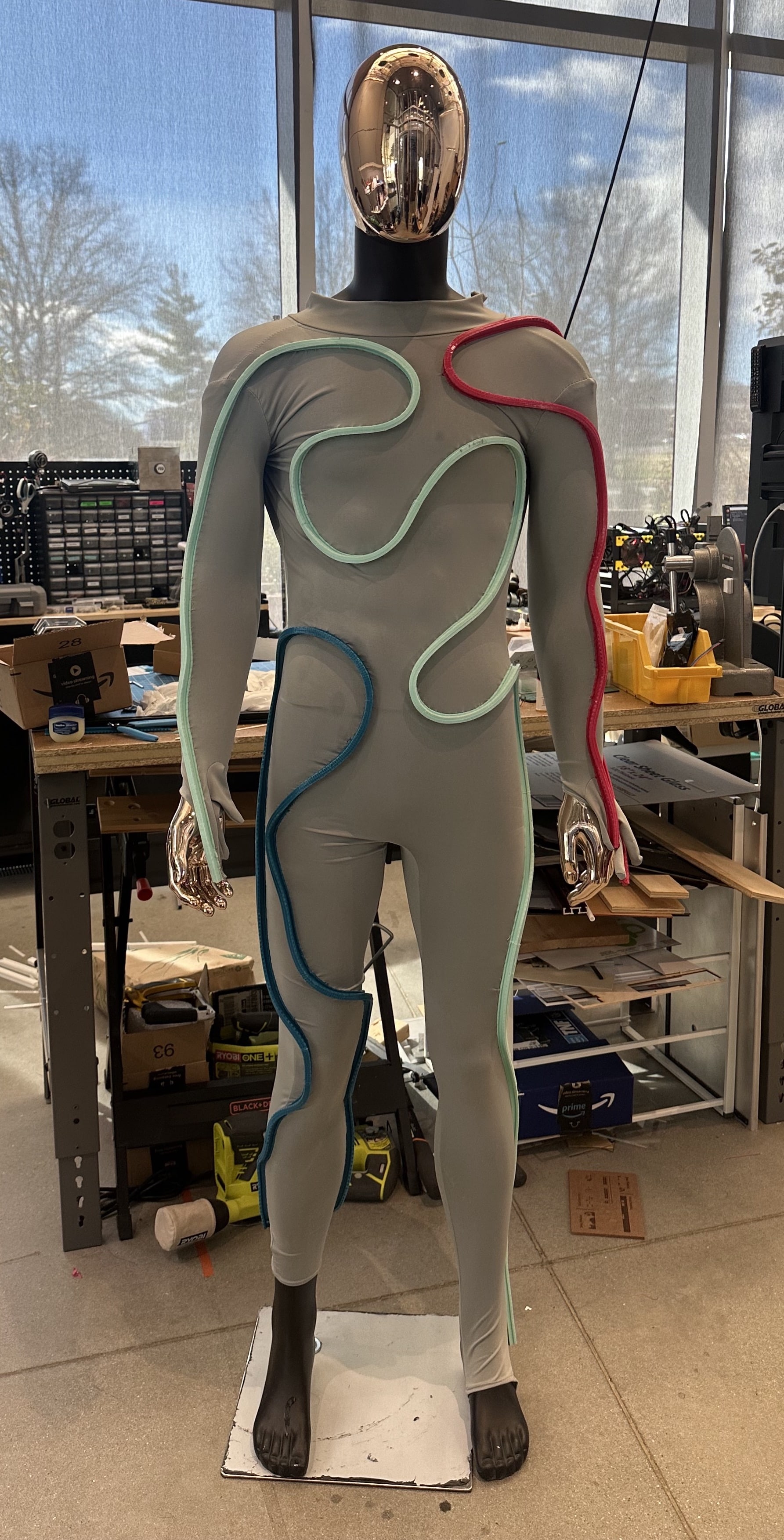}
    \caption{The performer's costume on a mannequin. Three pieces of silicone track is sewn onto the dancer's leotard. The dancer can move the robot between different pieces of track.}
    \Description{A grey leotard containing 3 separate winding tracks in dark blue, sky blue and red. The grey leotard is worn by a mannequin.}
    \label{fig:costume}
\end{figure}

\paragraph{The Moving Wearable Robot}

The wearable robot used in the performance builds on the Calico platform~\cite{sathya_calico_2022, antony2025minimal}, an open-source wearable robot designed to move along a soft, fabric-mounted track. 
In our implementation, the robot travels on a silicone track stitched onto the dancer’s leotard, allowing it to remain secure even during vigorous movement. Compact in size (42mm $\times$ 32mm $\times$ 35mm), the robot moves at approximately 15 cm/s and is wirelessly controlled by a backstage operator. 
It also includes RGB LEDs for displaying visual feedback during the performance.
The leotard features three distinct track segments (Figure~\ref{fig:costume}), each corresponding to a different section of the choreography. 
This arrangement invites the dancer to transition the robot across zones, creating moments of physical and narrative shift. 
While the hardware design evolved iteratively over the course of rehearsals, we focus here on the final configuration as used in the live performance.

\paragraph{Audience Interface and Control System}

\begin{figure}[t]
    \centering
    \includegraphics[width=1\linewidth]{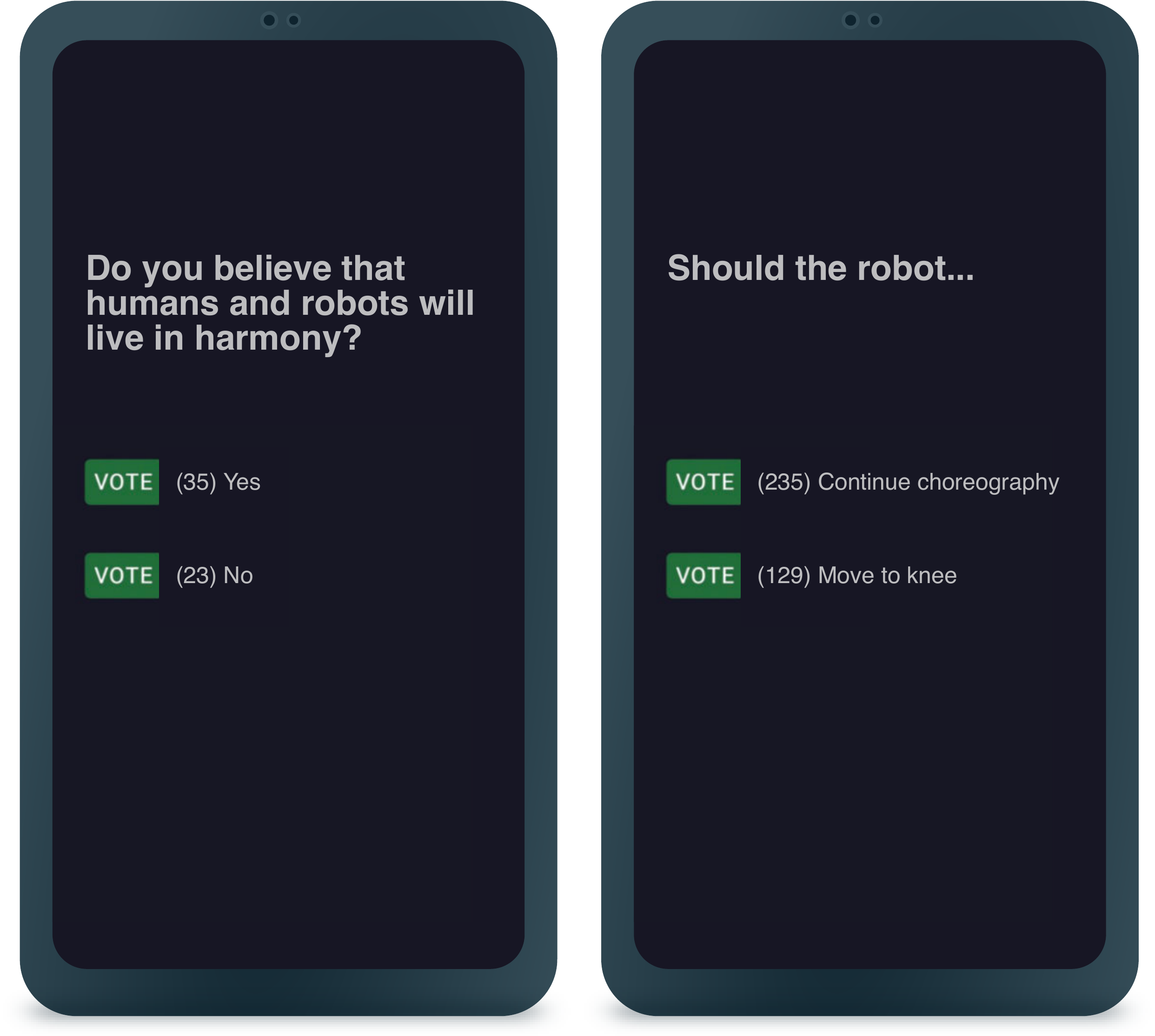}
    \caption{The voting interface used by the audience on a smartphone.}
    \Description{Two screens showing the voting interface. The left screen contains a prompt saying 'Do you believe that humans and robots will live in harmony?' with two options 'yes' or 'no'. 'yes' has 35 votes and 'no' has 23 votes. The right screen contains a prompt saying 'Should the robot...' with two options: 'Continue choreography' and 'Move to knee'.}
    \label{fig:interface}
\end{figure}

Audience members participate via their mobile phones using a custom web interface (Figure~\ref{fig:interface}). Prompts appear throughout the performance, allowing the audience to vote on the robot’s movement. 
These choices shape how the robot responds—and by extension, how the dancer interprets those movements in real time.
The voting interface is built with React and runs on a lightweight HTTP server written in Go, using BoltDB for temporary data storage. 
RESTful API endpoints enable seamless communication between components. 
This setup allows audience members to join or leave at any point during the performance without registration or setup, minimizing barriers to participation.
The robot is controlled by a backstage operator through a browser-based dashboard that communicates with the robot via Web Bluetooth. 
A Python script bridges the interaction system with Isadora, the performance software used for visuals and cue management. 
This script converts OSC (Open Sound Control) messages from Isadora into API calls and streams real-time voting data back to the stage, enabling live visualization of audience choices.

\paragraph{Privacy and Data Considerations} \label{section:privacy}

The system was intentionally designed to foreground collective participation while preserving audience privacy. 
No personal data or device identifiers are collected. 
Each vote is treated as an anonymous contribution to a collective decision, and no logs or session data are stored. 
By abstracting away individual identities, we create space for fluid participation and focus on the dynamics of collective agency. 
After each performance, anonymized vote data is stored locally in JSON format to support post-performance reflection and analysis.

\section{The Iterative Design of \dancetwo} \label{section:performance-design}

\begin{figure*}[t]
    \centering
    \includegraphics[width=\textwidth]{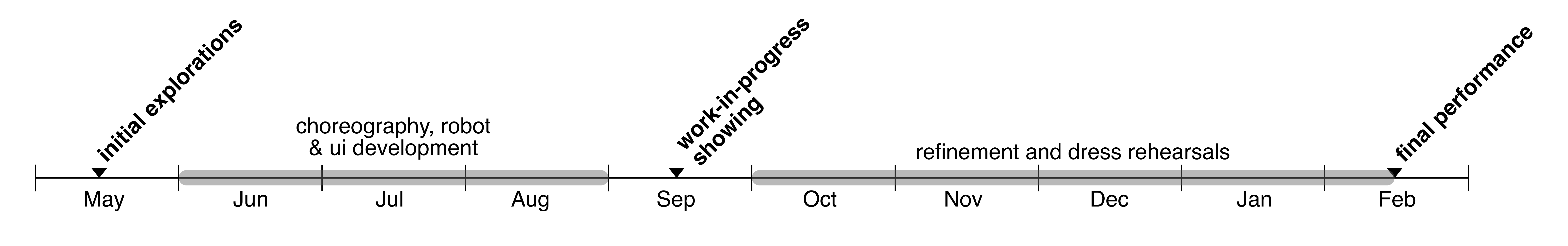}
    \caption{A ten-month timeline of the performance-led research agenda.}
    \Description{A timeline showing the development of the performance over 10 months starting from initial explorations, moving to choreography, robot & UI development, a work-in-progress showing, refinement and dress rehearsals and the final showing}
    \label{fig:timeline}
\end{figure*}

The development of \dancetwo unfolded over a ten-month long, performance-led research process (Figure~\ref{fig:timeline}). 
This process brought together ten interdisciplinary collaborators, including four leads responsible for robotic design, interaction and software development, choreography, and creative direction (IDs 1---4, Table~\ref{table:interviewees}). 
All four are co-authors of this paper and contributed firsthand reflections to this account.
Rather than starting from a fixed script or technical specification, the performance emerged iteratively through embodied exploration, studio prototyping, and reflection. 
From the outset, our goal was not merely to design a piece of interactive technology, but to explore how technology could mediate and redistribute agency across performer, audience, and machine. 
This approach resonates with both performance studies' emphasis on relational, emergent agency and HCI's growing attention to designing for collective and participatory experiences~\cite{schaffner_understanding_2022}.

\subsection{Initial Explorations And Early Concepts (Month 1)} The project began with a three-day workshop designed to build trust among team members and to explore the potential of wearable robotic movement in relation to dance. 
None of the four core leads had collaborated before, and the initial focus was on establishing a shared vocabulary across our domains.
The first day of the workshop was grounded in contact improvisation exercises~\cite{bodyresearch}, emphasizing weight sharing, guided touch, and fluid transitions. 
For the technical lead, who had no prior dance experience, this embodied activity provided an unexpected but effective entry point into the performative and collaborative ethos of the project. 
In retrospect, this mode of mutual grounding---through movement, rather than discourse---set the tone for the project's ongoing negotiation of artistic, technical, and conceptual priorities.

\begin{table*}[t]
    \caption{Anonymized descriptions of the leads in the design and performance team. All members are part of a large public institution in North America.}
    \centering
    \begin{tabular}{ r  p{1.9in}  p{2.4in}  p{2in}} 
     \hline
      & \textbf{Role} & \textbf{Background} & \textbf{Experience} \\
     \hline
     ID1 & Robotics Design and Pilot & Researcher \& Hardware Designer & 15+ years of hardware design\\
     ID2 & Choreographer and Systems Designer & Systems Designer and Amateur Modern Dancer & 20+ years of software and UI design \\
     ID3 & Choreographer & Dance Professor & 20+ years of dance\\
     ID4 & Creative Lead & Creative Director \& Producer & 15+ years of performance art\\ 
     \hline
    \end{tabular}
    \label{table:interviewees}
\end{table*}

\begin{figure}[b]
    \centering
    \includegraphics[width=1\linewidth]{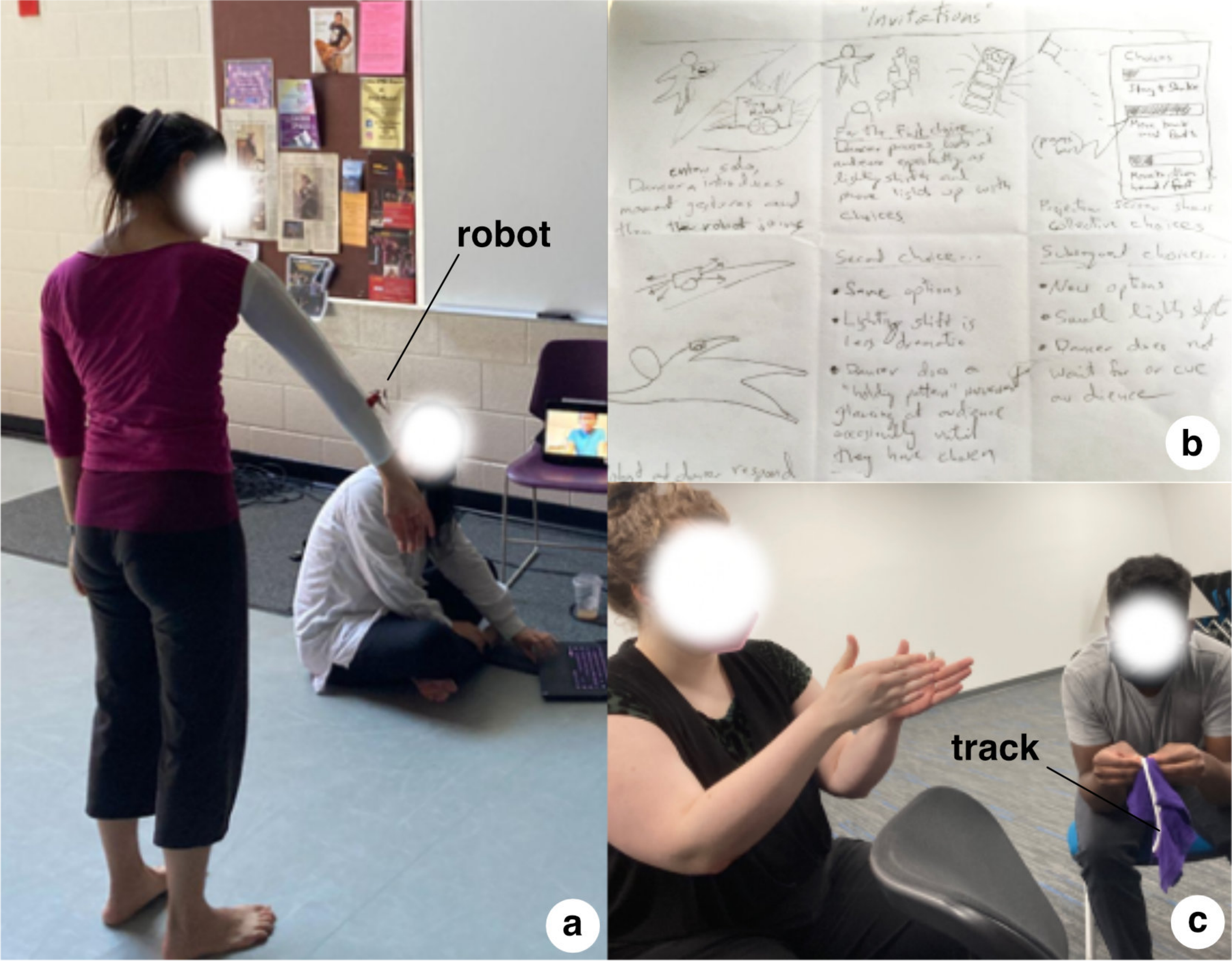}
    \caption{Early exploration and development of the \dancetwo performance. a) The choreographer understanding the capabilities of the moving wearable for the first time. b) Early brainstorming notes on engaging the audience through voting from their phones. c) Working with the costume designer on the garment.}
    \Description{Showing early explorations with the robot. A dancer is wearing a sleeve with the robot on it. There is an image of a whiteboard showing the software architecture and 2 people are sitting with the robot and track in hand, in deep discussion}
    \label{fig:early}
\end{figure}

Later that day, the team experimented with a prototype of the Calico robot~\cite{sathya_calico_2022}, a wearable robot that moves along a silicone track. 
Team members wore a sleeve-mounted track while the robot was piloted along their arms (Figure~\ref{fig:early}a). 
Though technically limited to bidirectional motion, the tactile quality of the robot's movement---especially when juxtaposed with intentional gestures like arm raises or blocks---evoked a sense of partnership. 
This moment of co-presence between dancer and robot became the first articulation of a ``duet logic'': a symbiotic relationship where control, response, and intention flow between the body of the performer and the robot.

On the second day, the team brought the robot into public and semi-public environments to explore how context shaped perception. 
These early trials drew attention and curiosity from passersby---suggesting that participation or even observation could be charged with ethical and aesthetic weight. 
These informal encounters reinforced our emerging interest in treating the audience not merely as spectators, but as active participants in the unfolding negotiation of agency.

On the final day, the team began synthesizing these threads. While early discussions had focused on the audience shaping the narrative or influencing high-level aspects of the performance, our explorations led us to a more provocative possibility: what if the audience could ``interfere'' in the performance---not by influencing the narrative, but by channeling their agency through the robot? 
In this triadic system, the audience influences the robot, the robot affects the dancer, and the dancer's responses recursively shape how the audience chooses to act. 
This layered feedback loop resonated with Breel's distinction between agentive behavior and the experience of agency~\cite{breel2025meaningful}, and reframed interactivity not simply as control, but as interference---a relational disturbance, charged with both tension and possibility.

\subsection{Early Performance Design (Month 2---4)}\label{node}
Over the next three months, individual sub-teams pursued parallel tracks: software development, costume and hardware prototyping, and framing. 
While technical and artistic progress advanced independently, frequent syncs ensured mutual alignment.
One key challenge was how to structure the audience's role.
Unlike typical interactive systems in HCI, where individuals control devices directly, our system had to translate collective input into embodied, real-time consequences within a live performance. 
Drawing on frameworks from immersive and interactive theatre~\cite{breel2022facilitating}, we designed a five-part performance structure that gradually scaffolded the audience's engagement---from observation, to agentive input, to open-ended reflection.

We adopted a repetitive-but-evolving movement motif, first introduced in the solo and duet phases (Parts 2---3). 
This allowed the audience to become familiar with the movement vocabulary before being invited to intervene. 
By the time collective input was introduced in Part 4, participants could draw on memory to evaluate the consequences of their choices---an approach similar to narrative scaffolding in \textit{The Money}~\cite{breel2022facilitating}. 
We introduced deliberate pauses at key moments, allowing the dancer to ``listen'' to the audience's choice before proceeding. 
These moments highlighted not only audience agency, but also the limits and consequences of that agency.

Another major decision was to project live voting data onto the stage (Figure~\ref{fig:teaser}, right). 
This real-time visualization amplified the visibility of collective agency, turning abstract input into a shared performance element. 
In doing so, it surfaced the performativity of choice itself---mirroring the feedback loop described by Fischer-Lichte~\cite{fischer2008transformative}, in which each action transforms the performance context.

\subsection{Work-in-Progress Showing (Month 5)}\label{section:unlimited}
At the five-month mark, we presented a condensed, four-part version of the performance to a small group of dance students and faculty. 
This showing served as a critical checkpoint for testing not only the technical integration of the voting system but also how the choreography and interaction design shaped the audience's understanding of agency. 
The performance was followed by a post-show discussion and informal reflection session with the attendees.

The audience feedback revealed two key insights. 
First, participants strongly identified the relationship between the dancer and the robot as ``symbiotic,'' echoing the core conceptual language that had emerged within our team. 
While the term agency was not used explicitly, the audience consistently described their sense of control, co-dependence, and mutual influence. 
One participant remarked:
\textit{``Sometimes it felt like the robot was controlling the dancer, then I realized that we---the audience---are the robot.''}

This comment captured the layered entanglement of agentic forces in the performance. 
Although audience members were not physically present on stage, they nonetheless experienced themselves as part of the choreographic system---mediating their influence through the robot. 
This speaks directly to Breel's distinction between agentive behavior and the experience of agency~\cite{breel2025meaningful}. 
While the audience's agentive input (i.e., casting votes) was simple and discrete, the felt consequences of those inputs were relational and expressive. 
Many participants interpreted the robot's responses as direct reflections of their will, even in cases where system constraints or choreography mediated the outcome. 
This gap---between visible action and perceived authorship---became a central design concern in the remaining development of the performance.

Second, a technical glitch led to a serendipitous discovery. 
Due to a misconfiguration in the voting system, audience members were able to cast unlimited votes instead of the planned single vote per prompt. 
Rather than breaking the interaction model, this created an unexpected surge of energy and engagement. 
Audience members began tapping furiously---some with multiple fingers---attempting to overwhelm the system in favor of their chosen option. 
Laughter, competition, and vocal expressions of excitement filled the room. 
The act of voting became performative in its own right.

What initially seemed like a flaw revealed something more profound: a new form of augmented agentive behavior, where participants attempted to scale their influence beyond the intended bounds of the system. 
Although the underlying logic remained one of collective decision-making, individual participants began to experiment with ways to amplify their voice—mirroring dynamics observed in online platforms where collective outcomes can be gamed or manipulated~\cite{del2016spreading, berghel2018malice}. 
This mechanic, while chaotic, surfaced a more complex reading of collective agency: one that was not merely cooperative, but competitive, playful, and emergent.

Importantly, this behavior also shifted the experience of agency. 
Some participants reported feeling more powerful and engaged once they discovered the possibility of voting multiple times, while others felt overwhelmed or uncertain about whether their individual input had any meaningful effect. 
This variability underscored a core insight from interactive performance studies~\cite{breel2022facilitating, salter2016indeterminate}: that participation alone does not guarantee empowerment. 
For agency to feel meaningful, participants must recognize and believe in the impact of their actions---whether or not that impact is actually realized.

Rather than fixing the ``bug,'' we chose to embrace it. By preserving the unlimited voting mechanic, we intentionally introduced ambiguity into the audience's relationship with the system. 
This ambiguity highlighted the messiness of collective authorship---a quality that aligned closely with our thematic goals. 
Just as the dancer must adapt to a robot she cannot fully control, the audience too must navigate a system in which their influence is partial, relational, and shaped by the actions of others.

\subsection{Refinement, Dress Rehearsals, and the Final Performance (Months 6 to 10)} 
In the final phase, we refined the choreography, finalized the costumes, developed lighting and projection cues, and composed original music. 
A fifth and final section was added to invite the audience into a more reflective, affective space---framing their earlier interactions not just as inputs, but as expressions of belief, desire, and uncertainty.

The performance premiered across four nights at a local performing arts venue, as part of a curated showcase of contemporary dance. 
Each evening featured a mixed bill, with \dancetwo concluding the program. 
Across four shows, we reached approximately 200 audience members. 
Exit surveys and performance data [$\S$~\ref{section:findings}] offered valuable insight into how participants interpreted their role, and how the technology shaped their experience of agency.

In sum, the iterative design of \dancetwo was not simply a matter of refining interaction techniques, but of staging a live, unfolding inquiry into how agency---individual and collective, human and machinic---is negotiated in real time.
As Alaoui and colleagues note~\cite{fdili_alaoui_making_2019}, performance-led design is inherently messy. Our contribution lies not just in the final artifact, but in the process of navigating this mess.

\section{Findings}\label{section:findings}

In this section, we present findings from two sources: (1) self-reported audience feedback gathered through a post-performance Likert-scale questionnaire and open-ended responses, and (2) voting data logged during the interactive segments of the live performance. 
Together, these datasets shed light on the disconnect that can arise between agentive behavior, the subjective experience of agency, and the extent of the audience's actual influence over the performance.
They also help surface how system design, choreography, and interaction framing shaped the relational dynamics between the dancer, the robot, and the collective audience. 
Rather than aiming for generalizability, our findings offer a situated, exploratory account of how agency is staged, felt, and shared in this specific performative context.

Unlike controlled lab studies where researchers can collect data in tightly regulated environments, research through public performance is inherently more fluid, unpredictable, and situated~\cite{benford_performance-led_2013, fdili_alaoui_making_2019}. 
Audience members primarily attend for the artistic experience, not to participate in research protocols---posing challenges to systematic data collection. 
Nonetheless, across four performances with over 200 attendees, we received voluntary survey responses from 150 participants. 
These self-reported insights, alongside voting data captured during the interactive portions of the performance, offer a unique lens into how audiences perceived, enacted, and shared agency during \dancetwo.

\subsection{Perceived Agency and Connection}

\begin{figure}[t]
    \centering
    \includegraphics[width=1\linewidth]{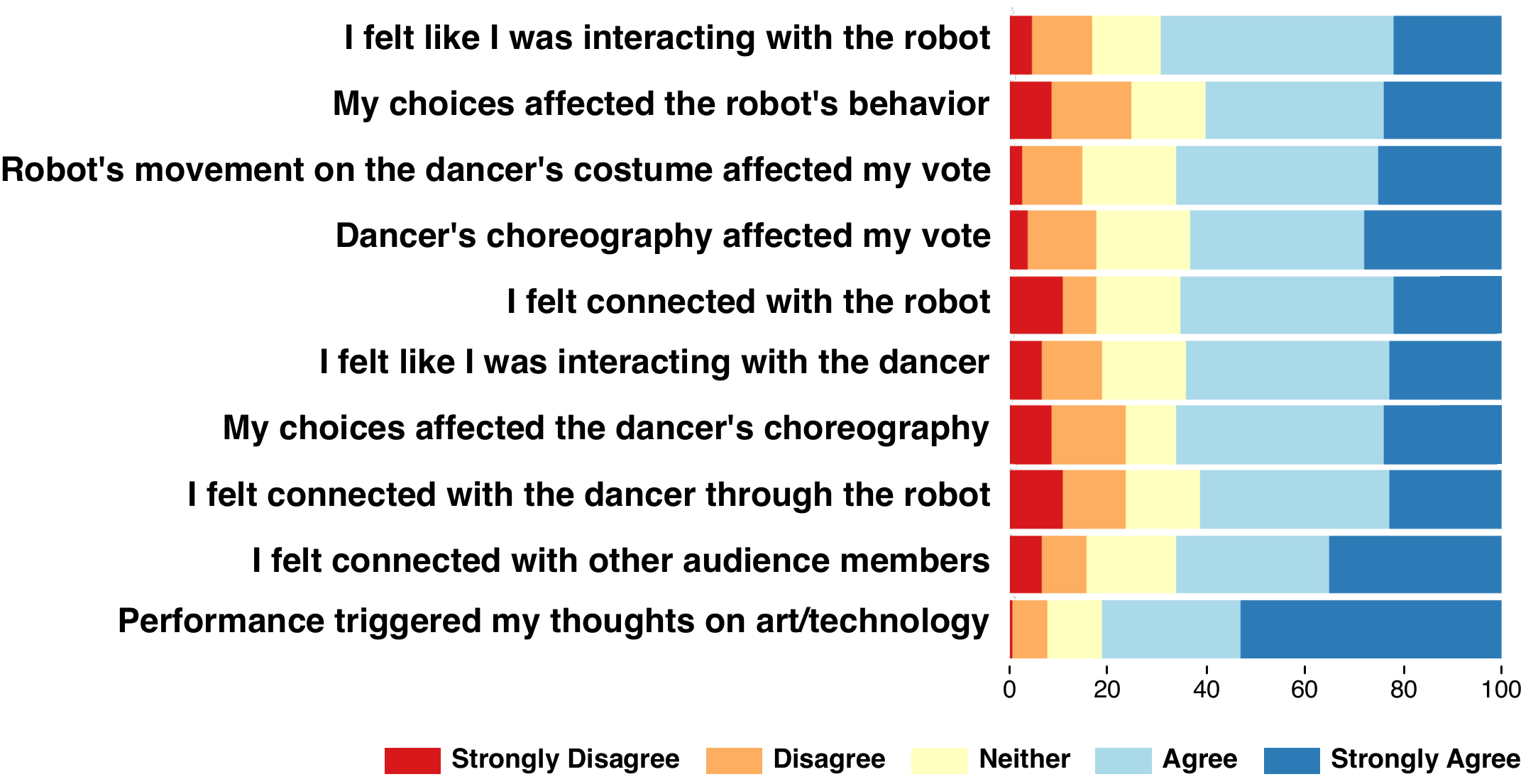}
    \caption{Self-reported audience responses regarding the perceived connections, affection, and relationships among the triad of audience, robot, and dancer. (150 volunteered responses in total.)}
    \Description{Showing likert scale responses from the audience survey. Participants responded to multiple questions: I felt that I was interacting with the robot, my choices affected the robot behavior, robot's movement on dancer's costume affected my vote, dancer's choreography affected my vote, I felt connected with the robot, I felt I was interacting with the dancer, My choices affect the dancer's choreography, I felt connected with the dancer through the robot, I felt connected with other audience members and performance triggered my thoughts on art/technology. Participants mostly somewhat agree or strongly agree on all prompts.}
    \label{fig:likert-scale}
\end{figure}

Figure~\ref{fig:likert-scale} presents results from a post-performance 10-question Likert-scale questionnaire. Overall, participants reported a strong sense of engagement with the piece. Notably, 81\% of respondents agreed that the performance prompted them to reflect on the relationship between art and technology—suggesting the choreography and interaction design succeeded in provoking thought beyond aesthetic appreciation.

Roughly 70\% of participants agreed that they felt they were genuinely interacting with the robot, while 64–65\% felt emotionally or perceptually connected to either the robot or the dancer. These findings speak directly to the audience's subjective experience of agency~\cite{breel2025meaningful}—the internal feeling of making an impact—even when their agentive behavior (e.g., voting) was minimal or mediated. In performance studies terms, this highlights the “autopoietic feedback loop”~\cite{fischer2008transformative} wherein meaning, action, and affect emerge dynamically between performers, systems, and observers.

Additionally, audience responses indicated that agency was not experienced in isolation. 
Around two-thirds of respondents reported that their choices during the performance were influenced by real-time cues---either the robot's behavior, visualized through its movement and LEDs, or the dancer's interpretative responses to the robot. This underscores the relational nature of agency in \dancetwo: rather than acting independently, audience members responded to the unfolding interaction between human and machine. 
These dynamics resonate with prior work in interactive and immersive theater~\cite{breel2022facilitating}, where action is often shaped by ambient cues and co-performance, even when explicit narrative control is limited.

Finally, one of the most notable findings was that 67\% of respondents reported feeling connected to other audience members (36\% strongly agreed, 31\% somewhat agreed)---marking the second-highest “Strongly Agree” rate across all questions.
While the system offered no direct peer-to-peer communication, the shared act of voting and witnessing its outcome in real time cultivated a sense of collective agency~\cite{fdili_alaoui_strategies_2015}. 
The design of the performance---particularly the visualization of votes and synchronized moments of decision-making---helped transform a room of individuals into a collaborative, if loosely coordinated, collective.

\textbf{Relational Influence and Performative Cues. }
As anticipated, the dancer's choreography---particularly her pauses, reactions, and expressions of frustration—had a significant influence on the audience's voting behavior. 
This aligns with prior research on persuasive design in interactive systems~\cite{lukoff_how_2021, baughan_i_2022}, which shows that users are often guided by subtle design cues even when presented with free choice. 
In our case, the performance structure itself---through repetition, visual feedback, and emotional expression---shaped the audience's use of their perceived agency.

These results reinforce the idea that agency in participatory systems is not only about what actions are available, but how they are framed and interpreted. 
Even when interaction is minimal, the framing of that interaction---both technically and theatrically---profoundly affects whether participants feel their input matters~\cite{breel2025meaningful, salter2016indeterminate}. 
In \dancetwo, the carefully choreographed interplay between dancer, robot, and system cues cultivated a performance environment where audience members felt both invited and compelled to act---even as their control was partial and mediated.

\subsection{The Votes Are In} \label{section:outcome}

\begin{figure}[t]
    \centering
    \includegraphics[width=0.8\linewidth]{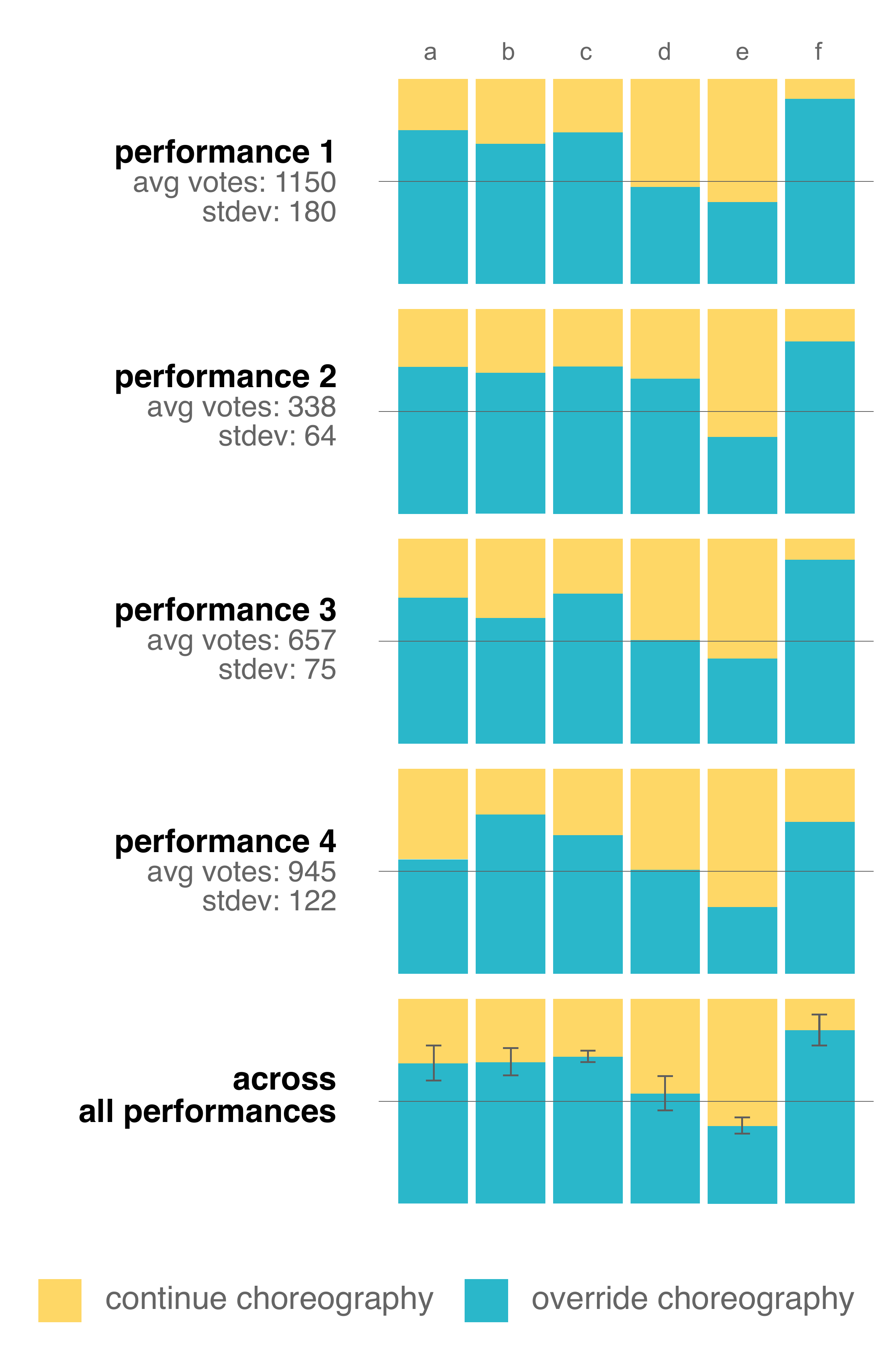}
    \caption{Synthesized voting data from Part 4 of all the performances. From top to bottom, we show a normalized ratio between votes to continue the choreography and votes to override the choreography. The top 4 charts show the data from each performance and the bottom chart shows consolidated data from across all performances. a-f indicates the 6 prompts on which the audience's votes would override the choreography.}
    \Description{5 stacked histograms are shown containing the vote share for each choice. The first 4 histograms show the data from each performance and the last histogram shows the vote share across all performances. The voting data between performances in largely similar. The votes to override choreography are higher in the beginning and they slowly continue to reduce until the last choice where the votes are suddenly overwhelmingly higher.}
    \label{fig:data}
\end{figure}

The audience's self-reported experience of agency becomes even more compelling when considered alongside the actual voting behavior recorded during the performances (Figure~\ref{fig:data}). 
Part 4 of the choreography featured six decision points (a---f) where the audience could collectively choose whether to let the robot continue along a predetermined path or override the choreography with an alternate behavior. 
Since participants were not required to vote---and were allowed to vote multiple times per prompt---we normalized the voting data to account for variability in total votes per performance and per prompt. 
This normalization enables us to examine not how often people voted, but how their collective choices distributed across the available options.

We focus specifically on Part 4 [$\S$\ref{section:part4}] because it represents the only phase in which audience choices directly altered the unfolding choreography. 
Parts 1---3 served to scaffold participation and establish expectations, while Part 5 emphasized reflection rather than redirection. 
Thus, Part 4 was the clearest site for analyzing how the audience enacted their agency---individually and collectively---within a constrained but dynamic system.

\begin{table*}[h!]
    \caption{The choices in Part 4 that allow the audience to override the choreography. $\mu$ is the mean ratio of audience that voted to override the choreography across all 4 performances and $\sigma/\mu$ highlights the variation in relation to the mean across all 4 performances.}
    \centering
    \begin{tabular}{ c  l  l  l  l } 
     \hline
    & \textbf{Continue Choreography} & \textbf{Override Choreography} & \textbf{$\mu$ (Override)} & \textbf{$\sigma/\mu$} \\
    \hline
     a & Move to right arm & Blink & 0.684 & 0.108 \\ 
     b & Travel across and all the way to the belly & Reverse to shoulder  & 0.690 & 0.083 \\
     c & Change color & Move back and forth & 0.717 & 0.034 \\
     d & Travel to left arm & Backtrack  & 0.536 & 0.135 \\
     e & Travel to ankle & Move back and forth & 0.378 & 0.090 \\ 
     f & Exit at the ankle & Vibrate & 0.846 & 0.078 \\ 
     \hline
    \end{tabular}
    \label{table:1}
\end{table*}

The results were strikingly consistent across all four performances. 
As Table~\ref{table:1} shows, the audience exercised their ability to override the choreography most strongly at the beginning and end of Part 4 (prompts a---c, f), and appeared more restrained in the middle (d, e). 
This arc suggests a collective behavioral pattern: initial curiosity and exploration of their power, a gradual pullback as the dancer expresses resistance, and a final surge of control when the stakes are implicitly raised---e.g., the suggestion that the performance might end unless they intervened.

These moments echo Breel's assertion that agentive behavior alone does not guarantee an experience of agency~\cite{breel2025meaningful}. 
The audience's actions were shaped not just by interface design but by the performative framing of each choice: the dancer's pauses, her gestures of struggle, and the emotive cues embedded in the lighting and music. 
The system did not simply ask, ``What do you want the robot to do?''—it staged a dilemma, inviting audiences to weigh their influence against the performer's autonomy.

Specific prompts further illustrate how collective decisions may have been shaped by narrative and emotional context. 
At prompt \textit{c}, where the override option was to make the robot move back and forth instead of changing color, the majority of the audience voted for override. 
This may reflect a desire to see a more active, kinetic response---a preference not uncommon in participatory experiences where visible impact correlates with a stronger sense of control. 
Yet at prompt \textit{e}, when the same override action (``move back and forth'') reappeared, the audience largely chose to let the choreography proceed. 
It's possible that repetition diminished novelty, or that by this point the audience felt more empathetic toward the dancer's increasing frustration---leading them to relinquish control.

While the dataset does not allow for definitive conclusions about audience motivation, it strongly suggests that perceived agency was shaped as much by the aesthetic and emotional design of the performance as by the mechanics of interaction.
The system carefully balanced control and ambiguity—designing for a structured indeterminacy that invited audiences to explore their influence without ever offering full control.

\subsection{Who Controls Whom?}

To conclude the survey, we included an open-ended prompt inviting general comments. 
Of the 150 respondents, 48 shared reflections. 
Many expressed appreciation for the piece and its thematic resonance, with comments such as:
\textit{``I really enjoyed this performance and watching humanity and machine learning interact''}
and
\textit{``Definitely poses lots of new thoughts and questions for artists/humans to think about.''}

One participant, however, offered a particularly insightful reflection—capturing the layered ambiguity at the heart of the piece:

\begin{quote} \textit{``I wondered throughout the performance if myself and other audience members were truly voting. Was the voting on my phone a simulation? Was the robot voting for me? I started thinking about this mostly because of the question about whether I controlled technology or technology controlled me. Even if the voting was controlled by the audience, the `choose your own adventure' format was compelling. I'm fearful of AI, but choosing our own adventure reminded me that technology can be positive so long as our choices are responsible, ethical, and democratic.''} \end{quote}

This reflection encapsulates the conceptual territory \dancetwo aimed to inhabit. 
Rather than offering certainty or resolution, the performance foregrounded relational ambiguity---a dynamic interplay of influence, response, and interpretation. 
By inviting the audience into a shared system of co-authorship, the work blurred the boundaries between control and surrender, individual intention and collective outcome. 
The question was never simply ``who controls whom,'' but rather, how is control felt, distributed, and made meaningful in a technologically mediated space of performance?

\section{Discussion} \label{section:discussion}

In this paper, we presented the design and implementation of \dancetwo, an interactive public dance performance that invites the audience to influence a wearable robot climbing on a dancer's body. 
Through a performance-led research process, we explored how interactive systems mediate individual and collective agency---not only by affording participation, but by shaping how participation is experienced. 
In this section, we frame our discussion through three intersecting theoretical lenses: (1) the distinction between agentive behavior and the experience of agency~\cite{breel2025meaningful}; (2) the relationship between perceived agency and actual power over a system~\cite{verbeek2005things}; and (3) how the choreography of agency in performative settings can reflect broader dynamics of human-technology interaction.

\subsection{From Agentive Behavior to the Experience of Agency}

The audience's experience of \dancetwo was characterized by a strong sense of agency. 
Survey responses show that participants felt they were genuinely interacting with the robot, influencing its behavior, and by extension, the dancer's movements. 
These reports highlight the strength of their experience of agency---a perception of authorship, even if their influence was distributed or mediated.
Mechanically, the system supported agentive behavior through real-time voting, visual feedback, and interaction prompts. 
The unlimited voting mechanic further amplified this sense of individual input by enabling participants to scale their effort and impact.

However, our analysis of the voting data complicates this perception. 
Across all four performances, voting patterns remained remarkably consistent, despite the open-ended structure. 
This suggests that while the audience felt empowered, their behavior was subtly guided—by the choreography, timing of prompts, visual framing, and emotional cues embedded in the performance. 
These constraints shaped the audience's actions without overtly restricting them---producing an illusion of open agency while steering collective outcomes. 
This echoes concerns in HCI about systems that support visible action but subtly limit choice~\cite{lukoff_how_2021, baughan_i_2022}, reinforcing Breel's argument that the experience of agency can be decoupled from its actual conditions.

\subsection{Perceived Agency vs. Actual Power}

This tension invites a deeper distinction between agency and power. 
While agency refers to the capacity to act intentionally within a system, power concerns the extent to which those actions meaningfully affect outcomes. 
In \dancetwo, the system afforded audience members moments of choice---but the structure of those choices, and the framing of when and how they appeared, significantly constrained their actual power over the performance. 
As our findings in $\S$\ref{section:outcome} suggest, key moments of audience engagement---such as the final vote to prevent the performance from ending---reflected emotional investment and perceived control, but also revealed how tightly the experience was choreographed.

This distinction mirrors dynamics in algorithmic systems more broadly, where users engage in visible behaviors (clicking, liking, swiping) and feel a sense of control, even as algorithms heavily determine the content they see or the outcomes they encounter~\cite{pariser2011filter, mathur2019dark}. 
Through this lens, the robot in \dancetwo acts not only as a performative partner but also as a proxy for contemporary technological systems---appearing responsive, while channeling interaction along pre-shaped paths. 
The audience is given the illusion of open-ended co-authorship, but their agency is staged, scaffolded, and ultimately bounded.

\subsection{Shifting Domains of Agency}

While voting prompts were framed around the robot's actions, the audience's influence often extended beyond that narrow scope. 
For instance, when given the option to ``end the performance,'' voting behavior surged---suggesting that the audience did not see themselves as simply controlling the robot, but as impacting the structure and trajectory of the entire performance.
In this way, perceived agency migrated from one domain (robot motion) to another (performance narrative), even though the system itself did not formally grant that level of control.

This echoes a broader theme in interactive systems: that agency is not always exercised where it is granted. 
As with social media platforms, users may begin by engaging with content but gradually assume broader roles---curating identities and shaping public discourse.
Similarly, the audience in \dancetwo began as choreographic participants but grew to see themselves as co-directors of the experience. 
As Reed Hastings famously put it, “We compete with sleep”~\cite{raphael_netflix_2017}---a reminder that influence can spill beyond system boundaries, even when those boundaries seem benign.

\subsection{Relational Framing and the Role of the Robot}

The robot in \dancetwo was deliberately designed to be small and visually non-threatening. 
It served as a bridge between the audience and the dancer---a mediator of agency. 
But this design choice also influenced how audience actions were interpreted. 
In a more dangerous or ambiguous context (e.g., Abramovi\'c's Rhythm 0~\cite{MarinaAbramovicRhythm}), audience agency might manifest as aggression or discomfort. 
In our case, the absence of risk or stakes may have encouraged participants to view their actions as harmless, even when they overrode the dancer's control.

This phenomenon highlights the ethical and aesthetic implications of interaction design. 
When systems feel safe, people may overestimate their benignity; when systems feel opaque or unpredictable, people may disengage or resist. 
In both cases, the experience of agency is shaped not just by mechanics, but by the relational framing of interaction~\cite{verbeek2005things}. 
The dancer's visible reactions, the robot's glowing LEDs, the beats the choreography---all shaped how the audience interpreted their own power.

\subsection{Data, Surveillance, and Role Reversal}

Finally, our decision to collect only aggregate, anonymous voting data reflected a commitment to audience privacy. 
However, this choice limited the granularity of insight we could obtain. 
Had we tracked individual behaviors---vote timing, frequency, shifts in response---we might have constructed more detailed portraits of how agency was distributed within the collective. 
Yet such tracking would have introduced new dynamics, potentially shifting the performance from a participatory experience into one of surveillance.

This raises a broader question about the ethics of data-driven design. 
In traditional performance, the audience watches and the performer is watched. 
In \dancetwo, the audience becomes visible---to the dancer, to the robot, and potentially to the system.
Had we chosen to collect more data, the roles of observer and observed could have inverted entirely, turning the performance into a live experiment in behavioral capture. 
As with algorithmic systems beyond the stage, the line between engagement and exploitation is thin, and easily crossed.

\section{Conclusion}

This paper explored the nuanced choreography of agency in \dancetwo, an interactive performance in which a live audience collectively influenced the behavior of a wearable robot through real-time voting. 
Framed through the lens of performance-led research, our work investigated how systems designed for participation shape not only what audiences do (agentive behavior) but also how they feel about their influence (experience of agency)---and whether that experience corresponds with actual power over the performance.

Our findings demonstrate that while participants reported a strong sense of engagement and perceived control, their collective behavior across fours performances followed consistent patterns---suggesting that their choices were subtly shaped by the system's design, choreography, and emotional framing. 
This reveals a critical tension between designed freedom and choreographed constraint, echoing broader concerns about how interactive technologies mediate perception, action, and agency.

By staging a triadic interaction between dancer, robot, and audience, \dancetwo surfaces key questions about authorship, influence, and the ethics of participatory systems. 
Audience members did not just interact with the robot; they engaged in a layered negotiation of control---sometimes amplifying their presence, at other times stepping back in recognition of the performer's autonomy. 
The performance thus becomes a lens through which to examine contemporary human-technology relations, where agency is distributed, relational, and often ambiguous.

Our contribution lies not only in the design of an interactive system or performance artifact, but in articulating a framework for analyzing the complex interplay between behavior, perception, and systemic power in participatory experiences. 
We offer this work as a provocation and methodological resource for designers, artists, and researchers exploring how interactivity can be used to both empower and critically unsettle assumptions about control in technologically mediated spaces.

\begin{acks}
This work was supported in part by the Arts for All initiative at the University of Maryland and by the National Science Foundation under Grant No. 2431575 and Grant No. 2229885 (NSF Institute for Trustworthy AI in Law and Society, TRAILS). Any opinions, findings, conclusions, or recommendations expressed in this material are those of the author(s) and do not necessarily reflect the views of the National Science Foundation. An LLM service was used for proofreading.
\end{acks}

\bibliographystyle{ACM-Reference-Format}
\bibliography{references}

\end{document}